\documentclass{pasa}%

\usepackage{graphicx}
\usepackage[amsmath,thmmarks]{ntheorem}
\usepackage{epstopdf}
\usepackage{setspace}
\usepackage{bbding}
\usepackage{amssymb}
\usepackage{gensymb}

\usepackage[T1]{fontenc}
\usepackage[utf8]{inputenc}

\newcommand{\spy}{s\,y$^{-1}$}
\newcommand{\Fig}{Fig.}
\newcommand{\cd}{c\,d$^{-1}$}
\newcommand{\kms}{km\,s$^{-1}$}
\newcommand{\eps}{erg\,s$^{-1}$}
\newcommand{\erg}{erg}

\title[Photometric Variability of the mCP Star CS\,Vir]{Photometric Variability of the mCP Star CS\,Vir: Evolution of the Rotation Period}

\author[D. Ozuyar et al.]{D. Ozuyar$^{1,2}$\thanks{dozuyar@ankara.edu.tr}~, H. T. Sener$^{3}$ and I.R. Stevens$^2$

\affil{$^1$Ankara University, Faculty of Science, Dept. of Astronomy and Space Sciences, 06100, Tandogan - Ankara / Turkey}
\affil{$^2$School of Physics and Astronomy, University of Birmingham, Edgbaston, Birmingham B15 2TT, UK}
\affil{$^3$Korea Astronomy and Space Science Institute, 776, Daedukdae-ro, Yuseong-gu, Daejeon, South Korea, 305-348}}

\jid{PASA}
\doi{10.1017/pas.\the\year.xxx}
\jyear{\the\year}

\usepackage{aas_macros}
\usepackage{hyperref} 

\hypersetup{colorlinks,citecolor=blue,linkcolor=blue,urlcolor=blue}

\hypersetup{draft}

\begin{document}

\begin{frontmatter}
\maketitle

\begin{abstract}
The aim of this study is to accurately calculate the rotational period of CS\,Vir by using {\sl STEREO} observations and investigate a possible period variation of the star with the help of all accessible data. The {\sl STEREO} data that cover five-year time interval between 2007 and 2011 are analyzed by means of the Lomb-Scargle and Phase Dispersion Minimization methods. In order to obtain a reliable rotation period and its error value, computational algorithms such as the Levenberg-Marquardt and Monte-Carlo simulation algorithms are applied to the data sets. Thus, the rotation period of CS\,Vir is improved to be 9.29572(12) days by using the five-year of combined data set. Also, the light elements are calculated as $HJD_\mathrm{max} = 2\,454\,715.975(11) +  9_{\cdot}^\mathrm{d}29572(12) \times E + 9_{\cdot}^\mathrm{d}78(1.13) \times 10^{-8} \times E^2$ by means of the extremum times derived from the {\sl STEREO} light curves and archives. Moreover, with this study, a period variation is revealed for the first time, and it is found that the period has lengthened by 0.66(8) \spy, equivalent to 66 seconds per century. Additionally, a time-scale for a possible spin-down is calculated around $\tau_\mathrm{SD} \sim 10^6$ yr. The differential rotation and magnetic braking are thought to be responsible of the mentioned rotational deceleration. It is deduced that the spin-down time-scale of the star is nearly three orders of magnitude shorter than its main-sequence lifetime ($\tau_\mathrm{MS} \sim 10^9$ yr). It is, in return, suggested that the process of increase in the period might be reversible.
\end{abstract}

\end{frontmatter}

\section{Introduction: Chemically Peculiar Stars}
\label{introduction}

Chemically Peculiar (CP) stars are located from the zero age main sequence (MS) to the terminal age MS. They have masses in the range between 1.5 M$_{\odot}$ and about 7 M$_{\odot}$ \citep{2014psce.conf...10N}, and their spectral types spread from early-B to late-F \citep{1996Ap&SS.237...77S}. Therefore, they contain stars with effective temperatures greater than 6\,500 K \citep{2005A&A...440L..37H}. In general, CP stars consist of A and B type peculiar (Ap and Bp, respectively) variables separated from normal stars which have the same effective temperature due to their abnormal chemical compositions and low rotational velocities (typically $v \sin i < 120$ \kms; \citealt{2000ApJ...544..933A}). The reasons of this peculiarity are the under-abundance of solar-like elements as well as over-abundance of metal and rare-earth elements seen in these stars \citep{2009arXiv0905.2565M}.

Based on their chemical peculiarity, CP variables are divided into four main subgroups as CP1, CP2, CP3 and CP4 \citep{1974ARA&A..12..257P}. Also, they are clearly split up into two main groups by the fact that CP2 and some of the CP4 stars show organised magnetic fields with a large-scale structure (typically from 300 G to about 30 kG), whereas CP1 and CP3 stars do not have such magnetic fields \citep{1974ARA&A..12..257P}. Although the production mechanism of these magnetic fields is still a matter of debate, {\sl the fossil field theory} is the most widely accepted mechanism which successfully explains the above mentioned field formation in Ap and Bp stars \citep{leb10}.

The formation of anomalous chemical compositions is another controversial issue for CP stars. Several hypotheses such as the interior nucleo-synthesis resulting from evolution \citep{1965ApJ...142..423F}, surface contamination of a normal star by a supernova companion \citep{1967PROE....6..145G}, radiative diffusion by gravity and radiation pressure \citep{1970ApJ...160..641M}, and the selective accretion of interstellar matter via the stellar magnetic field \citep{1971A&A....14....1H} have been proposed. However, none of these theories, except the radiative diffusion -- which leads to the accumulation or depreciation of the atoms at certain depths and causes some of the elements to become excessive at the surface -- can successfully account for the CP phenomenon \citep{2004IAUS..224..173M}.

Magnetic fields also play an important role in the distributions of elements at the stellar surface by directly affecting radiative acceleration and atomic diffusion. Such that the spherical symmetry of the diffusive segregation of chemical elements is broken
due to the Zeeman splitting and the Lorentz force. So, the diffusions that take place in the presence of horizontal and vertical magnetic fields differ from each other. This situation is thought to cause the formation of spots and rings of enhanced element abundance \citep{1981A&A...103..244M}.

With the combination of the stellar rotation, these non-uniformly distributed spot regions on the surface cause periodic variations in the average magnetic field characteristics, line profiles, spectral energy distribution, and brightness in different photometric bands \citep{2011IAUS..273..249K}, and all of these variations can best be explained by {\sl the oblique-rotator model} \citep{1950MNRAS.110..395S}.

From several observations, it is known that the phases of these periodic modulations are directly correlated \citep{1992AA...263..203C}. For some CP stars, the phase extrema of the variations coincide with each other (53\,Cam; \citealt{1960AcA....10...31J}), while some of the extrema occur in the anti-phase for some other stars (CS\,Vir; \citealt{2004IAUS..224..657M}). On the other hand, in some cases such as HD 83\,368 \citep{1999AstL...25..608P}, the extrema do not take place at the same time even though the periods of all these variations are same.

Most of the CP stars are slow rotators with periods roughly between a day and a week. Long-term observations have revealed that the geometry of the spots can remain stable on the surface for decades as a consequence of the slow rotation. Thereupon, surface distribution, rotational periods, and even rotational braking of some of these stars can be calculated with unprecedented accuracies \citep{2008A&A...485..585M, 2011IAUS..273..249K}.

Although a vast majority of the discovered CP stars do not exhibit a light curve or a period variation, it is reported that there is a small number of them such as SX\,Ari ($\dot{P}=0.02$ \spy; \citealt{2001A&A...375..982A}), V901\,Ori ($\dot{P} = 0.356$ \spy; \citealt{2014psce.conf..270M}), and CU\,Vir ($\dot{P}= 0.165$ \spy; \citealt{2013MNRAS.431.2106P}) whose rotation periods slightly or excessively change over decades, and it is thought that magnetic braking is the major cause of these variations.

The changes in period can be easily obtained through long-term monitoring of seasonal light curves. However, peculiarities of only a small fraction of CP have been investigated, and approximately one-tenth of these stars have been considered in terms of rotation or photometric periods. To determine these parameters, high-precision instruments and long-term observations are needed  (with an accuracy better than 0.005 mag; \citealt{2009arXiv0905.2565M}). 

In order to present the light variations and the evolution of rotational period of magnetic CP star CS\,Vir, this paper is organized as follows; in Sect.\,\ref{section2}, the literature studies of the star are reviewed; in Sect.\,\ref{section3}, the {\sl STEREO} satellite is briefly introduced; in Sect.\,\ref{section4}, the characteristics of the photometric data and the pipeline used for light curve analyses are described. In Sect.\,\ref{section5}, the results on CS\,Vir are presented, and finally the overall study is summarized and the results are discussed in Sect.\,\ref{section6}.

\begin{table}
\small
\begin{center}
\caption[]{Available period values for CS\,Vir. The periods, except the last four values, are around a mean value of 9.2954 days. The last four data excessively deviate from the mean value and do not fit the STEREO light curves very well.}
\begin{tabular}{c|l|l}\hline \hline          
\textbf{Time}	&	\textbf{Period}	&	\textbf{References}\\	
\textbf{(year)}	&	\textbf{(day)}	&	\\	
\hline
1944--1945	&	9.295	&	1 \\

1947--1950	&	9.2954	&	2\\

1949	&	9.295(3)	&	3\\

1953--1957	&	9.2954	&	4\\

1964	&	9.2957(2)	&	5\\

1964--1965	&	9.2954	&	6\\

1969	&	9.2954	&	7\\

1970	&	9.2954	&	8	\\

1969--1971	&	9.29477(5)	&	9\\

1970--1971	&	9.2954	&	10\\

1972--1974	&	9.2954	&	11	\\

1974	&	9.29541(7)	&	12	\\

1950--2000	&	9.29545(3)	&	13	\\

1989--1991	&	9.29571(18)	&	14	\\

1999	&	9.29545	&	15	\\

2007--2011	&	9.29572(12)	&	This study			\\

1969--2016	&	9.29558(6)	&	16	\\
\hline
1972--1973	&	9.3	&	17	\\

1990--1993	&	9.2918	&	18	\\

1990--1993	&	9.287	&	19	\\

2007	&	9.3105(13)	&	20	\\
\hline  \hline                      
\end{tabular}
\label{tab:table1}
\end{center}
\tabnote{\textbf{References.} (1)~\citet{1947ApJ...105..283D};
(2)~\citet{1951ApJ...114....1B}; (3)~\citet{1950MNRAS.110..395S}; (4)~\citet{1960stat.book..282B};
(5)~\citet{1965Obs....85..204A}; (6)~\citet{1969MNRAS.142..543H}; (7)~\citet{1970AA.....7...10M};
(8)~\citet{1971AJ.....76..422W}; (9)~\citet{1978AAS...31..205B}; (10)~\citet{1972AA....16..385M};
(11)~\citet{1985AAS...59..369P}; (12)~\citet{P75}; (13)~\citet{2004IAUS..224..657M}; (14)~\citet{1992AA...263..203C};
(15)~\citet{2001AA...365..118L}; (16)~\citet{2016AA...588A.138R}; (17)~\cite{1975PASP...87..221P}; 
(18)~\citet{2011MNRAS.414.2602D}; (19)~\citet{1997AA...323L..49P}; (20)~\citet{2012MNRAS.420..757W}.
}
\end{table}

\section{Literature Review of CS\,Vir }
\label{section2}

CS\,Vir (\hbox{HD 125\,248}; \hbox{HIP 69\,929}; HR 5355) (A9SrEuCr, $V = 5.86$ mag) is a well-known, bright magnetic CP (mCP) star that has attracted the attention of many researchers due to its line strength and light curve variabilities for several decades (Table~\ref{tab:table1}). 

\citet{1931ApJ....74...24M} was the first investigator who detected variable line intensities similar to those in \hbox{$\alpha^2$ CVn}. \citet{1947ApJ...105..283D} confirmed non-harmonic intensity variation and found that Eu II and Cr II lines changed in the same period ($\sim 9.295$ days), but in opposite phases (the maximum of Eu II modulation coincided with the minimum of Cr II). 

\citet{1950MNRAS.110..395S} photoelectrically obtained the light curve of the star with a period of 9.295 days ($A = 0.053$ mag), and found that the phase of the light maximum with respect to the Eu II variation was 0.55, but coincided with the maximum of Cr II variation. They also proposed the oblique rotator model for the first time in order to explain the photometric, spectral, and magnetic field
variations.

\citet{1951ApJ...114....1B} studied on the magnetic variability of CS\,Vir and explored a large reversing magnetic field with a variation period of around 9.3 days. They observed that the Eu II maximum occurred when magnetic polarity reached the positive maximum. Based on the variabilities in the observations, they stated that the star was a component of a binary system.

The spectroscopic observations by \citet{1969MNRAS.142..543H} verified that the star was a spectroscopic binary with the period of 4.4 years. On the basis of magnetic field observations, they showed that the rotational period of CS\,Vir was 9.2954 days. Unlike other studies, \citet{1978AAS...31..205B} found a slightly shorter period of 9.29477 days after three years observation, and reported that the light variation in V-band was double-waved whereas U- and B-bands showed single-wave curves.

Contrary to the arguments that Ap magnetic stars do not have a large inhomogeneities of oxygen, \citet{1992A&A...256L..31M} revealed that the atmosphere of CS\,Vir exhibited a strong and large inhomogeneous oxygen distribution.  Additionally, \citet{1992AA...263..203C} observed some variabilities in infra-red region and derived the best period to be 9.29571(18) days, compatible with visible light, spectrum and magnetic field variations.

\citet{2004IAUS..224..657M} improved the rotational period of star as 9.295450(30) days by combining 592 observations over 43 years. Finally, \citet{2016AA...588A.138R} obtained high-resolution spectropolarimetric observations of CS\,Vir to have a better understanding of the mechanism of atomic diffusion in the presence of magnetic fields. They constructed detailed maps of the surface magnetic field and abundance distributions for the star and showed that  its magnetic field has mostly been poloidal and quasi-dipolar with two large spots of different polarity and field strength. Combining 47 years of longitudinal field measurements with their own observations, they improved the rotational period of the star to be $P_\mathrm{rot} = 9.29558(6)$ days.

\section{The Solar TErrestrial RElations Observatory}
\label{section3}

The Solar TErrestrial RElations Observatory, {\sl STEREO}, is the third mission of the  \lq Solar Terrestrial Probes\rq\ program of the NASA. Two identically designed spacecrafts are positioned in a heliocentric orbit at radii of  \hbox{$\sim$ 1 au}. {\sl STEREO-A} (orbiting ahead of the Earth) and {\sl STEREO-B} (orbiting behind the Earth) orbit around the Sun while they drift away from the Earth in opposite directions. 

{\sl STEREO} monitors coronal mass ejections of the Sun and their propagations in the interplanetary medium. For this task, the satellites have been equipped with several instrument packages. The Sun Earth Connection Coronal and Heliospheric Investigation ({\sl SECCHI}) is one of these packages and includes the {\sl  Heliospheric Imagers (HI)}, which contain two visible-light cameras ({\sl HI-1} and {\sl HI-2}), referred to {\sl HI-1A}, {\sl HI-1B}, {\sl HI-2A}, and {\sl HI-2B}, depending on the satellite on which they are located. They produce photometric data by pointing near to the solar disk and monitoring brightness of background stars around the ecliptic ($V$ = 12 mag or brighter). The {\sl HI-1} instrument observes the stars in $20^{\circ}$ by $20^{\circ}$ field of view (FOV) with 40 minutes cadence for $\sim 20$ days while {\sl HI-2} has a $70^{\circ}$ by $70^{\circ}$ FOV and two-hour cadence. For details of the {\sl HI} instruments refer to \citet{2009SoPh..254..387E} and \citet{2010SoPh..264..433B}.

As periods of most mCP stars vary from several hours to days, {\sl STEREO} satellite is quite suitable to detect these periodic signals. Also, spectral energy distribution of mCP stars decreases with increasing wavelengths in visual spectral range. Since the spectral response of the {\sl HI} instruments is very broad (400 nm up to 950 nm), the window around 400 nm in the filter allows {\sl STEREO} to be sensitive to the variations of mCP stars \citep{2007A&AT...26...63M}.

For this study, only the data from {\sl HI-1A} are used since the data from the other three instruments are progressively deteriorating. A more detailed description of the basic light curves can be found in \citet{2011MNRAS.418.1325S} and \citet{2013MNRAS.431.3456W}.

\section{Photometric Data and Light Curve Analysis}
\label{section4}

Seasonal data comprise an observation interval of $\sim 20$ days. The cadence of the data is a photometry point every 40 minutes. This data set allows to perform analysis in a wide frequency range with the Nyquist frequency of around 18 \cd\ ($\sim 1.5$ h).

\begin{figure}
\begin{center}
\hbox{\hspace{0.55cm}\includegraphics[width=0.40\textwidth]{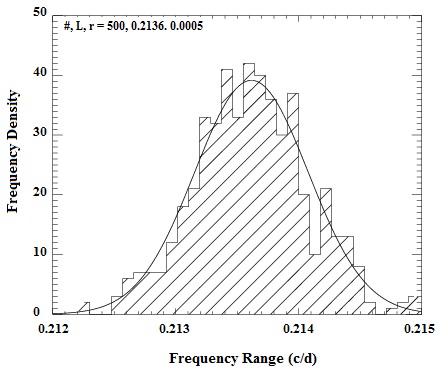}}
\caption[]{Frequency probability distribution of the data taken in 2007 and the best Gaussian fit made to calculate the most accurate frequency. \#, L and r denote the number of random frequencies used for histogram, frequency of cycles in a day and its error derived from the Gaussian fit, respectively.}
\label{fig:Figure 2}
\end{center}
\end{figure}

In order to analyse the data of CS\,Vir, the light curves are needed to be decontaminated from internal and external effects caused by the circumstances mentioned by \citet{2011MNRAS.418.1325S}. Therefore, long-term variations are removed from the light curves by using a $3^\mathrm{rd}$ order polynomial fit, which is the CCD response function of {\sl HI-1A}, and observation points greater than $3\sigma$ are clipped. Thus, the light curves cleaned from spurious effects are obtained.

\begin{table*}
\small
\begin{center}                                              
\caption[]{All available extremum times for CS\,Vir. In the first column, the archival time values (Raw Extremum Times) are given as Julian Date (JD). The star symbols indicate that these data have been interpolated from archival data sets. The obelisk symbols represent minimum times derived either from the light curves or from line intensities. Phase corrected extremum times, which have also been converted to Heliocentric JD, are presented in the fourth column. The archival data with no standard deviation value are marked with the asterisk symbols. The error values of these data are assumed to be $\pm 0.1$ days. Last two columns are the epoch and O-C values calculated from the maximum times.}
\begin{tabular}{l|l|l|l|r|r}\hline \hline    
\textbf{Raw Extremum Times}	&	\textbf{Reference}&	\textbf{Type}	&	\textbf{Corrected Maximum Times}	&\textbf{Epoch}&	\textbf{O-C}\\
\textbf{(JD)}	&		&			&\textbf{(HJD)}&&	(day)	\\
\hline	
	
2\,425\,309.6704$\star$$\dagger$	&	1	&	Line Intensity	&	2\,425\,314.7860	$\pm$	0.1000	&	$-$3163	&	1.1625	\\
2\,426\,378.6021$\star$$\dagger$	&	1	&	Line Intensity	&	2\,426\,383.7152	$\pm$	0.1000	&	$-$3048	&	1.0843	\\
2\,430\,133.7700$\star$$\dagger$	&	1	&	Line Intensity	&	2\,430\,138.8877	$\pm$	0.1000	&	$-$2644	&	0.7872	\\
2\,430\,143.0700$\dagger$	&	1	&	Line Intensity	&	2\,430\,148.1873	$\pm$	0.1000	&	$-$2643	&	0.7911	\\
2\,430\,384.7412$\star$$\dagger$	&	1	&	Line Intensity	&	2\,430\,389.8531	$\pm$	0.1000	&	$-$2617	&	0.7684	\\
2\,430\,905.2607$\star$$\dagger$	&	1	&	Line Intensity	&	2\,430\,910.3759	$\pm$	0.1000	&	$-$2561	&	0.7310	\\
2\,431\,936.9860$\star$$\dagger$	&	1	&	Line Intensity	&	2\,431\,942.1040	$\pm$	0.1000	&	$-$2450	&	0.6346	\\
2\,433\,103.9500	&	2	&	Light Max.	&	2\,433\,103.9523	$\pm$	0.1500	&	$-$2325	&	0.5183	\\
2\,440\,284.6800$\dagger$	&	3 &	Light Min.	&	2\,440\,280.0395	$\pm$	0.0100	&	$-$1553	&	0.3123	\\
2\,440\,372.9530$\star$	&	4	&	Light Max.	&	2\,440\,372.9579	$\pm$	0.1000*	&	$-$1543	&	0.2736	\\
2\,440\,373.0035	&	5$\dagger\dagger$	&	Light Max.	&	2\,440\,373.0000	$\pm$	0.0454	&	$-$1543	&	0.3157	\\
2\,440\,382.2500	&	4	&	Light Max.	&	2\,440\,382.2544	$\pm$	0.1000*	&	$-$1542	&	0.2743	\\
2\,440\,391.5464$\star$	&	4	&	Light Max.	&	2\,440\,391.5501	$\pm$	0.1000*	&	$-$1541	&	0.2744	\\
2\,440\,391.6467	&	5$\dagger\dagger$		&	Light Max.	&	2\,440\,391.6420	$\pm$	0.0522	&	$-$1541	&	0.3662	\\
2\,440\,405.4557$\dagger$	&	5$\dagger\dagger$		&	Light Min.	&	2\,440\,400.8027	$\pm$	0.0482	&	$-$1540	&	0.2312	\\
2\,440\,400.8408$\star$	&	4	&	Light Max.	&	2\,440\,400.8438	$\pm$	0.1000*	&	$-$1540	&	0.2723	\\
2\,440\,410.1396$\star$	&	4	&	Light Max.	&	2\,440\,410.1341	$\pm$	0.1000*	&	$-$1539	&	0.2669	\\
2\,440\,614.6407	&	5$\dagger\dagger$		&	Light Max.	&	2\,440\,614.6464	$\pm$	0.0590	&	$-$1517	&	0.2734	\\
2\,440\,679.7004	&	5$\dagger\dagger$		&	Light Max.	&	2\,440\,679.7024	$\pm$	0.0528	&	$-$1510	&	0.2594	\\
2\,440\,679.7072$\star$	&	6	&	Light Max.	&	2\,440\,679.7124	$\pm$	0.1000*	&	$-$1510	&	0.2694	\\
2\,440\,689.0038$\star$	&	6	&	Light Max.	&	2\,440\,689.0093	$\pm$	0.1000*	&	$-$1509	&	0.2706	\\
2\,440\,698.2891$\star$	&	6	&	Light Max.	&	2\,440\,698.2948	$\pm$	0.1000*	&	$-$1508	&	0.2604	\\
2\,440\,698.3130	&	7	&	Light Max.	&	2\,440\,698.3187	$\pm$	0.0120	&	$-$1508	&	0.2843	\\
2\,440\,707.5742$\star$	&	6	&	Light Max.	&	2\,440\,707.5800	$\pm$	0.1000*	&	$-$1507	&	0.2498	\\
2\,440\,726.1813$\star$	&	6	&	Light Max.	&	2\,440\,726.1867	$\pm$	0.1000*	&	$-$1505	&	0.2651	\\
2\,441\,051.5236$\star$	&	6	&	Light Max.	&	2\,441\,051.5291	$\pm$	0.1000*	&	$-$1470	&	0.2574	\\
	&	8	&	Light Max.	&	2\,454\,372.0510	$\pm$	0.0182	&	$-$37	&	0.0175	\\
	&	8	&	Light Max.	&	2\,454\,715.9750	$\pm$	0.0114	&	0	&	0.0000	\\
	&	8	&	Light Max.	&	2\,455\,069.2099	$\pm$	0.0341	&	38	&	$-$0.0023	\\
	&	8	&	Light Max.	&	2\,455\,403.8410	$\pm$	0.0231	&	74	&	$-$0.0170	\\
	&	8	&	Light Max.	&	2\,455\,757.0882	$\pm$	0.0199	&	112	&	$-$0.0071	\\

\hline  \hline                      
\end{tabular}
\label{tab:table3}
\tabnote{\textbf{Notes.} $\dagger\dagger$ http://astro.physics.muni.cz/mcpod/}
\tabnote{\textbf{References.} (1)~\citet{1947ApJ...105..283D};
(2)~\citet{1950MNRAS.110..395S}; (3)~\citet{1978AAS...31..205B}; (4)~\citet{1970AA.....7...10M}; (5)~mCPod; (6)~\citet{1972AA....16..385M}; (7)~\citet{2004IAUS..224..657M}; (8)~STEREO (this study).}
\end{center}
\end{table*} 

The light curve of CS\,Vir has presented a sinusoidal characteristic due to a spot modulation on stellar surface. All analyses are therefore performed using the Lomb-Scargle (LS) algorithm since this method is sensitive to such variations \citep{1976Ap&SS..39..447L, 1982ApJ...263..835S}. During the analyses, the number of independent frequencies ($N_\mathrm{id}$) is calculated by employing $N_\mathrm{raw}/2$, where $N_\mathrm{raw}$ is the number of observation points in the raw data (e.g. $N_\mathrm{id}= 325$ for light curve of CS\,Vir taken in 2007). Also, false alarm probability (FAP) is assumed to be 99\% ($P_\mathrm{0} = 0.01$). Apart from these, signals are sought between the frequency range of 0.05 -- 18 \cd, and variabilities greater than the Nyquist frequency are not taken into account. 

\begin{table*}
\small
\begin{center}
\caption{Details of the seasonal and combined observations of CS\,Vir. Observation years, observation lengths, start and mid times of each observation, the numbers of raw and cleaned data points, frequencies and amplitudes derived from the seasonal and combined data sets with the Lomb-Scargle method are given in the table. Due to some satellite-related problems such as pointing discontinuity or tracking error, the numbers of seasonal raw data points are less than 720 points (maximum points derived in 20 days).}
\begin{tabular}{l@{\hskip 0.12in}c@{\hskip 0.12in}c@{\hskip 0.12in}c@{\hskip 0.12in}c@{\hskip 0.12in}c@{\hskip 0.12in}r@{\hskip 0.12in}r@{\hskip 0.12in}}\hline \hline
\textbf{Time} &\textbf{Data Length}&\textbf{Obs. Start Time}&\textbf{Mid-Obs. Time}&\textbf{Data Pnt.}&\textbf{Data Pnt.}&\textbf{Freq.}	&\textbf{Amp.}\\

\textbf{(year)} &\textbf{(day)}&\textbf{(HJD)}&\textbf{(HJD)}&\textbf{(raw \#)}&\textbf{(cleaned \#)}&\textbf{(\cd)}&\textbf{(mmag)}\\
\hline									
2007	& 19	&	2\,454\,363.9729	&	2\,454\,373.5979&	650	&	638	&	0.1068(2)	&	11.13(19)	\\
2008	& 18	&	2\,454\,709.3063	&	2\,454\,718.1396&	656	&	587	&	0.1056(2)	&	15.09(16)	\\
2009	& 19	&	2\,455\,053.1118	&	2\,455\,062.6118&	252	&	249	&	0.1075(5)	&	12.05(46)	\\
2010	& 17	&	2\,455\,399.5007	&	2\,455\,408.1534&	605	&	605	&	0.1092(3)	&	12.46(22)	\\
2011	& 18	&	2\,455\,742.1118	&	2\,455\,751.3063&	673	&	633	&	0.1094(2)	&	16.04(21)	\\\hline
Combined	& 91	&	2\,454\,363.9729&	2\,455\,062.2368&      2836	&       2712    &	0.107576(1)	&	13.33(11)	\\
\hline  \hline                      
\end{tabular}
\label{tab:table2}
\end{center}
\end{table*}

The Levenberg-Marquardt Optimisation (LM-fit) Method is applied to the curve by defining a simple Fourier series to determine a model of the examined sinusoidal light curve. Moreover, more than one Fourier series with different coefficients and their derivatives are defined in the case of any other harmonics in the light curve. Frequency and amplitude values derived from the LS analysis of the cleaned light curve are used as initial parameters. As a result of 5\,000 iterations, the most accurate fit with the smallest error value is obtained.

Once a model light curve is derived, the most accurate frequency and its uncertainty are assessed using the Monte-Carlo simulation algorithm. To do this, random Gaussian noise with the mean of zero and the sigma value determined from the cleaned curve is produced and the noise component is then added to the model fit. Subsequently, another LS procedure is implemented to this noisy curve. Once a frequency value is identified, all these processes are repeated by adding another noise component to the model curve. After 500 repetitions, 500 random frequencies are produced around the actual frequency. When a probability distribution is performed by a histogram, a well-defined peak, close to the expected value, is plotted on the diagram, and the most accurate frequency is calculated using a simple Gaussian fit (\Fig~\ref{fig:Figure 2}). Moreover, the final frequency error is found from the standard deviation of these 500 frequencies.

\begin{figure*}
\begin{center}
{\includegraphics[width=\textwidth, height= 0.75\textheight]{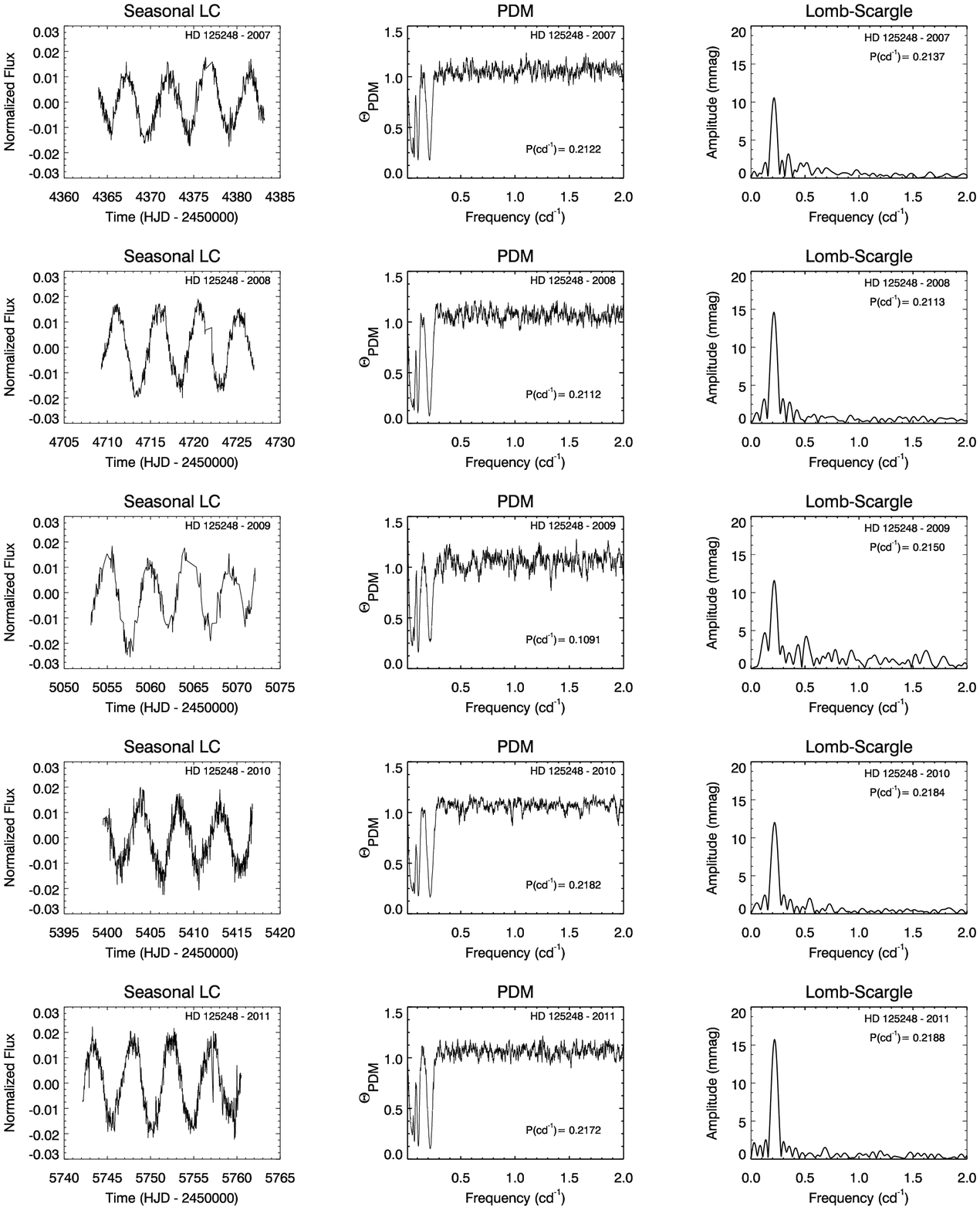}}
\caption[]{Annual light curves, related PDM and LS analyses of CS\,Vir are shown in the columns. In the first column, the x-axis of the light curves represents the observation duration in Heliocentric JD and the y-axis shows normalized flux count calculated from ($F(t)/\overline F) -1$.}
\label{fig:figure3}
\end{center}
\end{figure*}

All processes taking part in the analyses are repeated by using the Phase Dispersion Minimization (PDM) technique to make a comparison with LS frequencies. The results are also compared with the period values given in the literature and with data derived by the Hipparcos satellite. Phased light curve profile of the star is generated based on periods derived from combination of all seasonal light curves.

The maximum times are obtained from the seasonal  light curves by means of the Kwee-van Woerden method  \citep{1956BAN....12..327K} to investigate period variability over years, and those with the smallest errors are put together with data from the literature (Table~\ref{tab:table3}). In the first column of the Table~\ref{tab:table3}, the archival extremum times (Raw Extremum Times) are given as Julian Date (JD). The star symbols in this column indicate that these data have been interpolated either from line intensity measurements or photometric data sets. Also, the obelisk symbols ($\dagger$) in the same column represent minimum times derived either from the light curves or from line intensities at $\phi = 0.5$. These data have been subjected to phase correction for shifting to zero phase. Phase corrected extremum times have been converted to Heliocentric JD and are  presented in the fourth column of the Table~\ref{tab:table3}. The archival data with no standard deviation are marked with the asterisk symbols ($\star$). The error values of these data are assumed to be $\pm 0.1$ days. 

The times of photometric maximum light of the star are presented in the form of;
$$HJD_\mathrm{max} = T_\mathrm{0} + PE + \frac{1}{2}\frac{dP}{dt} \overline{P}E^2, $$ 
where $T_\mathrm{0}$ is the zero epoch, $P$ is the rotational period in days, $E$ is the number of cycles and $\frac{1}{2}\frac{dP}{dt} \overline{P}$ represents the long-term variation in period.

\section{Results}   
\label{section5}

In this study, five-year light curves of CS\,Vir taken between 2007 -- 2011 are analyzed, and the details related to the data are given in Table~\ref{tab:table2}. Due to some satellite-related problems such as pointing discontinuity or tracking error, the numbers of seasonal raw data are less than 720 points (the maximum number of observation points taken in 20 days). Except the light curve data taken in 2009, the curves consist of $\sim$ 600 data points. Even though the data sets do not have the maximum number of observation points, the sinusoidal structure of the light curve is clearly observed as shown in \Fig\ \ref{fig:figure3}. 

\begin{figure*}
\begin{center}
{\includegraphics[width=\textwidth, height= 0.75\textheight]{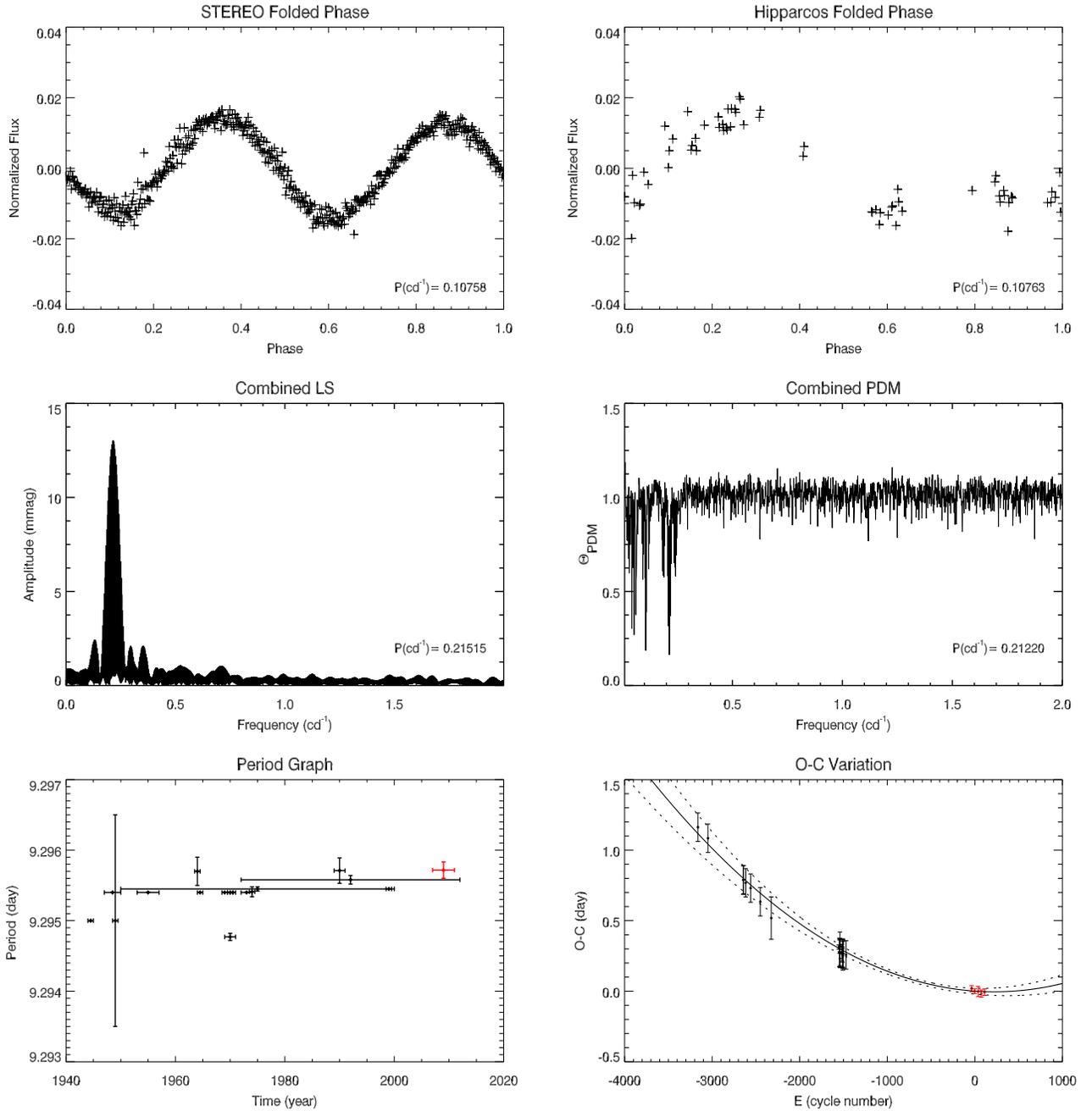}}
\caption[]{Analysis results of CS\,Vir. Folded light curves produced by the {\sl STEREO} and the {\sl Hipparcos} periods are given in the first row. The phase folded light curves do not have the same shape since the {\sl Hipparcos} observations have been performed using $Hp$ filter \citep{1997ESASP1200.....E}. A sample $Hp$ observation of the star can be also seen in \citet{2004IAUS..224..657M}. LS and PDM  analyses of the combined light curve are presented in the second row. Period and O-C variation graphs of CS\,Vir are shown in the last row. The long-term rotational change in the observed time
of the light maximum minus the calculated time of light maximum is expressed in days.}
\label{fig:figure4}
\end{center}
\end{figure*}

For the analyses, the LS method is initially applied to the seasonal data and a period value of around 4.64 days, half of the literature period, is acquired for each year. Following this, each data set is analyzed by using the PDM technique. As the number of light cycles is limited, the PDM method also detects a period of $\sim$ 4.60 days for the seasonal data sets (except for 2009). On the other hand, the frequency analyses of the five-year combined data show that the actual period is around 9.29 days. Therefore, the seasonal periods are assumed to be the twice of the LS and PDM results.

The LS and PDM periods of the five-year data combination are used to check general light curve characteristics of CS\,Vir. Since the LS period (0.107576(1) \cd) provides a more suitable folded curve having the less scattered data points and the smallest sigma value, it is chosen for the further calculations. As shown in \Fig\ \ref{fig:figure4} (upper left), the main light curve exhibits a double-waved structure that is not observed in most studies discussed in Sect.\,\ref{section2}. In this plot, it is seen that one of the maxima is slightly shallower than the other. Moreover, the analysis of the Hipparcos data produces a similar period (0.10763 \cd) to that of the {\sl STEREO} (\Fig\ \ref{fig:figure4}, upper right).

Thereto, a possible long-term period variation is investigated by collecting all available archival data, even though a stable period is reported by several authors. More than 20 periods have been given in archives as seen in Table~\ref{tab:table1}. The majority of these values are around a mean period of 9.2954 days, whereas the last four data in the table excessively deviates from this value and do not fit the {\sl STEREO} light curves very well. Therefore, they have not been used for our calculations. In \Fig\ \ref{fig:figure4} (left bottom), archival and combined {\sl STEREO} period values are shown with black and red diamond symbols, respectively. From the graph, it is not easy to deduce any period change in the star, because the vast majority of the data are gathered around the mean value. However, our five-year combined data suggest a moderate increase in the period if it is assessed with the result from \citet{2016AA...588A.138R}.

For the purpose of substantiating this period change, $E$ and O-C values are determined from the maximum times given in the Table~\ref{tab:table3} and presented in the last two columns of the same table. Data, covering a long time interval, come from
different observers using various detectors and different pass-bands. Such data are not homogeneous and their effects in the O-C diagram should be appropriately evaluated. Therefore, the errors of the O-C values are calculated by using the error propagation method. Accordingly, all errors from the maximum times $T$, reference maximum time $T_0$ and period $P$ are considered even if the uncertainties of $T_0$ and $P$ do not significantly affect the result. On the other hand, based on \citet{2005ASPC..335....3S}, the cycle number $E$ is assumed to be error-free since it has been a function of $T$. Together with the errors, O-C values are shown in \Fig~\ref{fig:figure4} (right bottom). As seen from the figure, the data display a parabolic variation, which opens upward, since 1930. To calculate the rate of this parabolic change, O-C values are fitted with the Levenberg-Marquardt least-squares method, which iteratively minimizes the sum of the squares of the errors between the data points and the function through a sequence of updates to parameter values. The solid and dashed lines in the figure are the best fit to the data and the 1-$\sigma$ uncertainty from the fit, respectively. As a result of this fitting procedure, the period variation is found to be around 0.66(8) \spy, or 66(8) seconds per century. 

Based on these results, the times of maximum light of the star are computed as: $HJD_\mathrm{max} = 2\,454\,715.975(11) +  9_{\cdot}^\mathrm{d}29572(12) \times E + 9_{\cdot}^\mathrm{d}78(1.13) \times 10^{-8} \times E^2$.

\begin{table*}
\begin{center}
\caption[Physical parameter of CS\,Vir]{Physical parameters of CS\,Vir. Temperature, luminosity and mass values have been adopted from \citet{2006A&A...450..763K}. Radius and rotational velocities were estimated from period, temperature, and luminosity values.}
\begin{tabular}{c|c|c|c|c|c|c}\hline \hline
HD	&	$\log (L/$L$_{\odot})$	&	$\log (T)$	&	$M$	&	$R$	&	$v_\mathrm{eq}$	&	$P$	\\
	&		&		&	(M$_{\odot}$)	&	(R$_{\odot}$)	&	(\kms)	&	 (day)	\\\hline
125\,248	&	1.50(8)	&	3.992(1)	&	2.27(7)	&	1.95(18)	&	11(1)	&	9.29572(12)	\\
\hline \hline
\end{tabular}
\label{tab:table4}
\end{center}
\end{table*} 

\section{Discussion and Summary}  
\label{section6} 

In this study, photometric data of CS\,Vir are derived from {\sl STEREO} satellite between 2007 and 2011 to investigate its period evolution. During the analyses, five years of seasonal light curves and a compilation of these data are examined with one of the most widely-used frequency detection methods, Lomb-Scargle. By cleaning the light curves from the distorted effects, the best periods are obtained with a precision of $10^{-4}$ and $10^{-6}$ \cd\ for the seasonal and combined data sets, respectively. These are then used to investigate a possible variation in period.  From the {\sl STEREO} light curves, the seasonal maximum times with the smallest error values are calculated and supplemented with all the relevant archived observations that are available to study the period evolution seen in the O-C diagram. 

\begin{table*}
\begin{center}
\caption[Period variation parameters]{Period and its variation rate as well as spin-down and main sequence lifetime of CS\,Vir.}
\begin{tabular}{c|c|c|c|c|c}\hline \hline
HD	&	$P$	&	$dP/dt$	&	$\dot{P}/P$	&	$\tau_\mathrm{SD}$	&	$\tau_\mathrm{MS}$\\
&	 (day)	&	(s~yr$^{-1}$)	&	(s$^{-1}$)	&	(yr)	&	(yr)	\\\hline
125\,248	&	9.29572(12)	&	0.66(8))	&	2.62(30)$\times 10^{-14}$	&	1.21(35)$\times 10^{6}$	&	1.29(9)$\times 10^{9}$	\\
\hline \hline
\end{tabular}
\label{tab:table5}
\end{center}
\end{table*}

Based on the examination of the {\sl STEREO} and archival data, it is detected that CS\,Vir has an explicit period variation in its O-C diagram. The data collected since 1930 indicated that the period has been gradually increasing over years. Accordingly, the rotation of CS\,Vir has slowed down by 0.66(8) seconds per year. Such a deceleration in rotation suggests a possible decrease in the kinetic energy of the star. Using the physical parameters given in Table~\ref{tab:table4}, the energy and the rate at which the energy decreased are roughly calculated as $E=1.02(19) \times 10^{45}$ \erg\ and $dE/dt = -5.33(1.17) \times 10^{31}$ \eps.  According to its period variation, the spin-down time-scale of the star is approximately $\tau_\mathrm{SD} = 1.21(35) \times 10^{6}$ yr. The main sequence lifetime of the star is also found as $\tau_\mathrm{MS} = 1.29(9) \times 10^{9}$ yr from the equation of $\tau_\mathrm{MS} = 10^{10}~\mathrm{yr} \times (M/M_{\odot})^{(1-\alpha)}$, where $\alpha = 3.5$ for main sequence stars and $10^{10}$~yr is the approximate lifetime of the Sun in the main sequence (\citealp{ghosh2007, koupelis2007quest, 1994sipp.book.....H}).

As stated by several researchers such as \citet{2009arXiv0905.2565M}, strict periodicity due to slow rotation is a common property of the majority of CP stars. However, it is reported in the literature that a small number of CP stars shows notable period changes within several years. CU\,Vir ($\dot{P}/P = 5.28 \times 10^{-14}$ s$^{-1}$) and V901\,Ori ($\dot{P}/P = 1.33 \times 10^{-13}$ s$^{-1}$), which have sinusoidal period variations, as well as BS\,Cir, which has a moderate rotation deceleration ($\dot{P}/P = 3.02 \times 10^{-14}$ s$^{-1}$), are some of the most important samples among these type of stars \citep{2014psce.conf..270M}.

In relation to these period variations, \citet{2006ASPC..355...27M} suggest that the moment of inertia and rotation period should change in mildly rotating stars showing no substantial angular momentum loss, and that evolution models in fact predict a slow down in their stellar rotation. \citet{2014psce.conf..270M} point out that the fastest change in period is seen in the most massive stars ($\tau_\mathrm{MS} \approx 30$ M\,yr), and is roughly $\dot{P}/P = 1.04 \times 10^{-15}$ s$^{-1}$. However, change rates given for CU\,Vir, V901\,Ori, and BS\,Cir are several times greater than evolutionary changes. From the findings in this study, such a result is also confirmed. The variation in period of CS\,Vir ($\dot{P}/P \sim 10^{-14}$ s$^{-1}$) is ten times greater than that of the most massive CP stars ($\dot{P}/P \sim 10^{-15}$ s$^{-1}$) discussed by \citet{2014psce.conf..270M}, and also quite compatible with their findings. In addition, the spin-down time-scale of the star is found around $\tau_\mathrm{MS} \sim 10^6$ yr. This value is nearly three orders of magnitude shorter than the MS lifetime of the star ($\tau_\mathrm{MS} = 10^9$ yr). This, in turn, suggests that the process of the increase in the period of CS\,Vir might be reversible. If so, the length of these cycles could be roughly estimated to be 248(14) yr for CS\,Vir.

It is not possible to measure rotational deceleration caused by stellar evolution of MS stars with current methods, since variation rates are quite small. Also, if rotational evolution is a consequence only of evolutionary changes, the rotational periods of CP stars would be relatively constant \citep{2014psce.conf..270M}. In spite of this, observational facts indicate that the periods of some CPs vary. As a result, numerous hypotheses related to the origin of these changes have been offered. These hypotheses, all of which assume that these stars rotate as a solid body \citep{1950MNRAS.110..395S}, can be outlined as; (1) mass and radius variations during MS evolution, (2) angular momentum loss because of standard and magnetized stellar winds, (3) precession of rotation axis, (4) light-time effect occurring due to an additional component.

The reason for the period decrease in CP stars can be explained by the first item given above. However, it has been already mentioned that detection of rotational slowing due to evolutionary reasons is not possible since variation is at least three times larger than the observational limits. Also, the spin-down time-scale caused by the standard stellar winds cannot be measured, because it is much larger than the evolutionary time-scale \citep{2014psce.conf..270M}. For the third explanation, the effects of precession in rotation axis should be seen as cyclic changes in light curves. Yet, there is no peculiarity observed in the {\sl STEREO} light curves of CS\,Vir, and no archival records have been found related to such changes. Also, \citet{2014psce.conf..270M} indicate that the amplitude of these variations originating from precession can be ignored. Finally, the last mechanism is a light-time effect, which causes a cyclic variations in the O-C diagram. Even though it has been reported that the star is a spectroscopic binary, this case is also not appropriate as its O-C variation shows a parabolic change.

On the other hand, this parabolic O-C diagram may be indicating a mass transfer existing between the components. If such a situation is indeed in question, the increment in the period should be the result of a mass transfer occurring towards the more massive component from the less massive one. When the semi-amplitude of the radial velocity variation ($K=7.60(17)$ \kms) and the orbital period ($P_\mathrm{orb}=1\,618(8.1)$ days) are adopted from \citet{1973ApJS...25..137A}, and $\sin^3i$ is assumed to be 0.679 \citep{1992Ap&SS.194..143H},  the mass of the companion star is estimated to be around 1.0 M$_{\odot}$. According to this, the radii of the Roche Lobes of the components are calculated as $RL_\mathrm{1} = 123$  R$_{\odot}$ and $RL_\mathrm{2} = 87$ R$_{\odot}$ by using the formula given by \citet{1983ApJ...268..368E}, and this result reveals that there is no mass transfer in this system yet.

Considering the options outlined above and the fact that period variation process might be reversible due to shorter  spin-down time-scale than that of MS lifetime, the rigid rotation hypothesis should be discarded and the differential rotation model should alternatively be discussed in detail as expressed by \citet{1998CoSka..27..205S}. As a result of such a rotation, an interaction between the layers takes place and hence, a cyclic increase and decrease in the moment of inertia occurs \citep{1998CoSka..27..205S}. This means that an unexpected alternating variability of rotation periods can be observed. In this case, rotation deceleration in CS\,Vir may be interpreted as a consequence of torsional oscillations produced by meridional circulations being in interaction with a magnetic field, and of rotational braking in outer layers caused by angular momentum loss via magnetically-confined stellar wind. Such explanation, which the angular momentum loss that occurs due to magnetic winds can be calculated for the hot CP stars with strong magnetic fields,  is also supported by \citet{2014psce.conf..270M}.  

There are numerous compiled catalogues with hundreds of CP stars, including \citet{2013MNRAS.429..119P} and \citet{2012MNRAS.420..757W} ({\sl STEREO}), \citet{2009A&A...506..213N} (CoRoT), and \citet{2008AstBu..63..139R} (ground-based). There are also several studies related to surface structures (CoRoT stars HD 49\,310 and HD 50\,773; \citealt{2015A&A...574A..57P, 2010A&A...509A..43L}), and pulsational behaviour of CP stars (CoRoT object HD 45\,975; \citealt{2014A&A...561A..35M}). However, few studies have been performed to achieve a better understanding of the period changes of these variables \citep[etc.]{2010A&A...511L...7M, 2014psce.conf..270M}, and almost none of them are based on space-based missions. Therefore, this investigation is quite important in terms of contributing to the studies on the period evolution of CP stars. 

To increase the accuracy and precision of the results, the amount of data collected between 2007 and 2011 can be increased. For this purpose, the data of the {\sl HI-1A} obtained after 2011 can be collected, or light curves observed by the {\sl HI-1B} and the {\sl HI-2} can be made compatible with {\sl HI-1A} data. Additionally, the FOVs of SMEI and K2 overlap the {\sl HI-1} image. Thus, if there is any available data provided by these missions, the current data can be combined with them. As a result, further information about the evolutionary stage of CS\,Vir can be obtained.

\section*{Acknowledgements}

We acknowledge assistance from Dr. Vino Sangaralingam and Dr. Gemma Whittaker
in the production of the data used in this study.

This work has made use of the BeSS database, operated at LESIA,
Observatoire de Meudon, France: http://basebe.obspm.fr

The STEREO Heliospheric imager was developed by a collaboration that
included the Rutherford Appleton Laboratory and the University of
Birmingham, both in the United Kingdom, and the Centre Spatial
de Lige (CSL), Belgium, and the US Naval Research Laboratory
(NRL),Washington DC, USA. The STEREO/SECCHI project is an
international consortium of the Naval Research Laboratory (USA),
Lockheed Martin Solar and Astrophysics Lab (USA), NASA God-
dard Space Flight Center (USA), Rutherford Appleton Labora-
tory (UK), University of Birmingham (UK), Max-Planck-Institut fr
Sonnen-systemforschung (Germany), Centre Spatial de Lige (Bel-
gium), Institut dOptique Thorique et Applique (France) and Institut
dAstrophysique Spatiale (France). This research has made use of
the SIMBAD data base, opened at CDS, Strasbourg, France. 
This research has also made use of NASA's Astrophysics Data System.



\end{document}